# Predicting Pulmonary Hypertension by Electrocardiograms Using Machine Learning

Eashan Kosaraju

Praveen Kumar Pandian Shanmuganathan

22 April 2023

**Abstract:**

Pulmonary hypertension (PH) is a condition of high blood pressure that affects the arteries in the lungs and the right side of the heart (Mayo Clinic, 2017). A mean pulmonary artery pressure greater than 25 mmHg is defined as Pulmonary hypertension. The estimated 5-year survival rate from the time of diagnosis of pulmonary hypertension is only 57% without therapy and patients with right heart failure only survive for approximately 1 year without treatment (Benza et al., 2012). Given the indolent nature of the disease, early detection of PH remains a challenge leading to delays in therapy. Echocardiography is currently used as a screening tool for diagnosing PH. However, electrocardiography (ECG), a more accessible, simple to use, and cost-effective tool compared to echocardiography, is less studied and explored for screening at-risk patients for PH. The goal of this project is to create a neural network model which can process an ECG signal and detect the presence of PH with a confidence probability. I created a dense neural network (DNN) model that has an accuracy of 98% over the available training sample. For future steps, the current model will be updated with a model suited for time-series data. To balance the dataset with proper training samples, I will generate additional data using data augmentation techniques. Through early and accurate detection of conditions such as PH, we widen the spectrum of innovation in detecting chronic life-threatening health conditions and reduce associated mortality and morbidity.

**Introduction:**

Pulmonary hypertension (PH) is a form of high blood pressure that affects the arteries in the lungs and the right side of the heart (Mayo Clinic, 2017). Pulmonary hypertension gradually worsens and can be life-threatening for some. With pulmonary hypertension, the blood vessels develop an increased amount of muscle in the wall of the blood vessels. The heart pumps blood from the right ventricle to the lungs where the blood gets oxygenated. Because the blood in the pulmonary arteries travels short distances, pulmonary artery pressures are typically low. Pulmonary hypertension is defined by a mean pulmonary artery pressure greater than 25 mmHg. When the pressure in the artery gets too high, the arteries in the lungs can narrow, and then the blood does not flow as well as it should due to increased resistance, resulting in pulmonary hypertension.

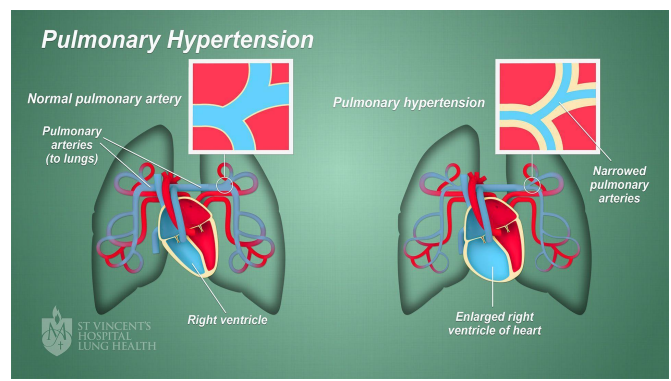

Figure 1: Diagram of Pulmonary Hypertension (Health, S. V. H. (n.d.).)

Although no proper cure exists for pulmonary hypertension, treatment can help reduce symptoms and improve quality of life. A wide array of treatments exist ranging from simple oxygen treatment and blood thinners(help prevent blood clots) to in the worst case lung and heart transplants (Mayo Clinic, 2017). Symptoms often develop slowly and can be unnoticeable for months and in certain scenarios the symptoms take years to manifest. Common symptoms include ankle and leg swelling, cyanosis (Bluish/Grayish discoloration of the skin), chest pain/pressure, dizziness, and a feeling of weakness. A wide variety of variations of PH exist. According to WHO, pulmonary hypertension is classified as Pulmonary Arterial Hypertension (PAH), PH due to Left Heart Disease, PH due to Lung Disease, PH due to chronic blood Clots (CTEPH), and finally PH due to miscellaneous causes (PHA Staff 2018). Pulmonary Hypertension has many causes, ranging from autoimmune diseases such as rheumatoid arthritis, obstructive sleep apnea, congenital or birth defects, left-sided heart failure, a range of medicines, and chronic lung disease. Additionally, PH can also lead to numerous side effects as the condition progresses. These include anemia, which is when there aren't enough healthy red blood cells to carry oxygen to the organs, arrhythmias, in other words, irregular heartbeats, blood clots, a potentially life-threatening complication when the blood in the veins or arteries when blood changes from liquid to partially solid, and finally pericardial effusion, which is an accumulation of fluid in the pericardium, or the sac that holds your heart.

      The Global Burden of Disease Study 2019 has identified PAH as a major contributor to right heart failure (Vos et al., 2020). The estimated 5-year survival rate from the time of diagnosis of pulmonary hypertension in the United States is only 57% without therapy and patients with right heart failure only survive for approximately 1 year without treatment (Benza et al., 2012). Pulmonary hypertension accounts for a substantial disease burden in both developed and developing countries (Gidwani & Nair, 2014). Unfortunately, 98% of the global PH burden is noted in resource-limited developing countries with an estimated 20-25 million people suffering from pulmonary hypertension, and another 64 million are at-risk and remaining undiagnosed (Butrous et al., 2008). Approximately, 21% of patients had symptoms for >2 years before diagnosis as per the REVEAL registry (Registry to Evaluate Early and Long-Term PAH Disease Management) (Benza et al., 2010).

      Given the indolent nature of the disease, early detection of pulmonary hypertension remained a challenge. Often, the diagnosis of pulmonary hypertension is made well after the disease process is set in or noted as an incidental finding during testing for other cardiac

conditions. The typical diagnostic modalities used to diagnose pulmonary hypertension include echocardiography, a non-invasive ultrasound modality used predominantly as a screening tool, and pulmonary artery catheterization, an invasive study used to confirm the diagnosis of pulmonary hypertension where a catheter is introduced into the right side of the heart and lung blood vessels to measure the pressures directly. Both echocardiography and invasive pulmonary artery catheterization are resource-driven modalities and need highly skilled personnel limiting their widespread use in resource-limited communities. On the other hand, a simple tool like an electrocardiogram with machine learning techniques can be used on a larger scale to screen and detect pulmonary hypertension with good accuracy. Pulmonary hypertension itself cannot be diagnosed by the ECG but numerous signs like right atrial enlargement, right atrial overload, and right ventricular hypertrophy suggest the presence of PH. The ECG changes are more evident as the disease progresses with changes to the right heart chambers, the right atrium, and the right ventricle.

     ECG is a widely accessible simple test that can be used as an initial screening tool if only the diagnostic accuracy can be improved. Using machine learning techniques, the subtle ECG changes early in the disease could potentially be identified. Early detection is often life-saving (Lau, E. M. T., Humbert, M., & Celermajer, D. S). In a study adjusting for age, sex, and PAH subtype, a delay of greater than two years was associated with an 11% increased risk of death (Weatherald, J., & Humbert, M). AI has been used for other conditions such as detecting Lung Cancer with CT scans and has proved to be a helpful tool for physicians. This same process can be carried over to a range of diseases, including Pulmonary Hypertension, a condition not yet extensively studied with AI models. Rising healthcare costs have become one of the biggest challenges in our time to deliver quality and equitable healthcare. AI can help bridge this gap by

making early disease detection more accessible to all populations, regardless of their socioeconomic status.

Going forward in our world, AI models will play a major role in medicine and the market is projected to reach over 300 Billion USD by 2026 (Kasyanau, A). It will put patients at significant risk if these models cannot accurately diagnose the disease condition and can be life-changing in rare scenarios. On the bright side, Artificial Intelligence is the future of medicine and will be a crucial tool that healthcare professionals can rely on. If implemented properly, AI is expected to save approximately 400,000 lives yearly (The socio-economic impact of AI in Healthcare, 2020).

Common ECG findings of pulmonary hypertension include right ventricular hypertrophy, right atrial abnormalities, and right axis deviation. ECG criteria used to determine the aforementioned abnormalities will be evaluated using a machine learning model. Right Ventricular Hypertrophy(RVH) is an enlargement or pathologic increase in muscle mass of the right ventricle in response to pressure overload, most commonly due to severe lung disease (Bhattacharya, P. T., & Ellison, M. B.). The function of the right ventricle (RV) is a major determinant of prognosis in pulmonary hypertension (Ryan, J. J., & Archer, S. L.).

In an electrocardiogram, the "p" wave reflects the atrial (the upper chamber) activity, and the "R" and "S" waves reflect the electrical stimulus as it passes through the ventricular walls (*The R Wave - Sinus Rhythm* ). The walls of the ventricles are very thick due to the amount of work they have to do and, consequently, more voltage is required. The S wave is the first downward deflection of the QRS complex that occurs after the R wave ("S Wave." ). However, an S wave may not be present in all ECG leads in a given patient ("S Wave."). Below in Figure 1, is an image that helps visualize the beats. The ECG leads provide a graphical description of the electrical activity of the heart (The ECG Leads). There are 12 such leads, which are placed at different parts of the body. The data used to train the ML model in this project will utilize all 12 leads. Finally, the criteria needed to diagnose RVH include Right Axis Deviation, an abnormal direction of depolarization of the ventricles represented by a QRS axis, a tall, and narrow R wave in lead V1, and lastly an R wave taller than the S wave in lead V1 ("Right Ventricular Hypertrophy and the ECG.").

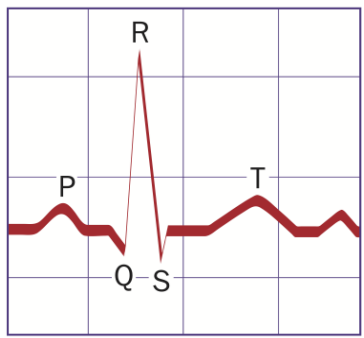

Figure 2: ECG Waves (Do et al., 2021)

The goal of this research project is to achieve accuracy better than the current research literature. Several procedures like Transfer Learning, Fine Tuning, and Domain Adaptation training will be used to improve the accuracy and overall performance of the model. When the model achieves the desired accuracy, an overarching front-end user-facing application will be created that can take in input ECG data, and display results. This application will process the user's ECG and run the model. An inference will be made from the output of the model. The output of the model will inform the probability of the user's ECG having RAO (Right Atrial Overload) or RVH (Right Ventricular Hypertrophy), both conditions which hint at the presence of pulmonary hypertension.

**Methodology:**

A literature search and review were conducted on the resultant electrocardiographic changes in pulmonary hypertension. Electrocardiographic studies (ECG) of patients with pulmonary hypertension appear to have changes reflecting right ventricular hypertrophy and/or right atrial overload. As a result, there was a need to find datasets that have electrocardiographic studies with findings of right ventricular hypertrophy and right atrial overload. Upon some research, an existing public dataset called PTB-XL was found, which fits the criteria necessary for this project. It had labeled ECG data indicating these conditions. The dataset had high and low-resolution samples of ECG labeled for different cardiac conditions. Python programming language was primarily used for preprocessing all the data from the dataset. The data fields available in the dataset were the age, sex, weight, and height of different users. Custom preprocessing scripts were written to ingest the binary-encoded ECG data into numerical arrays that can be fed into Machine Learning models for analysis and training down the line.

Initially for pre-processing, a standard normalization of all the features available in the dataset was conducted. A variety of graphs, including pie charts (all categories of data) and bar graphs(all categories of data), were plotted using Matplotlib to thoroughly understand the distribution of the data. Prior health conditions, described in the dataset, were defined to understand the procedures taken to diagnose patients involved in this data set. The data for the conditions of right atrial overload and right ventricular hypertrophy was then extracted from the PTB-XL dataset, using a Python library known as Pickle. This allowed the data to be saved in a binary encoded format which makes it easy to load and unload for different experiments in a simple and easy way rather than loading the entire dataset which happened to be very large. The filtered dataset was then split into a testing, training, and batch validation set for an unbiased training experience during the training process. A function from Scikit-Learn known as test_train_split was used for this. Test_train_split splits the samples into a training and testing category. 75% of the data was used for training and 25% of the available data was used for testing. This allowed for an organized dataset that can be tested upon different model architectures as needed.

In the existing literature, several model architectures were explored, and most of them were made available off the shelf using the library, TensorFlow. A pre-trained model was imported as a baseline. They then performed transfer learning to tweak the weights for the new model. This allowed a state-of-the-art model to be trained on this study's target domain with new data samples. The models will be trained on the training data and further tested for accuracy on the testing data. The target categories include right ventricular hypertrophy and right atrial enlargement, also known as right atrial overload. An important part of model training is to recognize overfitting, a vital issue in many models. Overfitting happens when the model, instead of learning, memorizes the training data and has very bad performance when tested on the validation set. Overfitting can be solved through a number of methods, including plotting graphs or using Keras's EarlyStopping program. Training will include testing various models of neural networks to find which one most accurately predicts Pulmonary Hypertension. As the model trains, the Learning Rate Decay API from Keras will be utilized to exponentially reduce the learning rate as the model starts approaching a plateau. Accuracy metrics including the train test losses, ROC curves, and F1 scores will be generated using the sci-kit-learn library. This process will be repeated for multiple models in order to find a model which attains an accuracy, which is desirable and safe for the patient.

**Results:**

The dense neural network (DNN) (Figure 3) achieved an accuracy of 98% on the training portion of the dataset. 50 Epochs (Epoch: when an entire dataset completes one cycle of forward and back propagation) were completed in the training process. Accuracy ranged between 97% and 99% during most epochs (Figure 4).

In addition, the model was tested on validation data. Validation data is a section of data that the model has never seen before. Training on validation data is a crucial part of model validation, as the model could memorize the training data. The validation data, which is new to the model, can help with detecting overfitting, a phenomenon when the model represents a 'fake' accuracy by memorizing instead of training on data. When testing on the validation data, the accuracy was between 97% and 99% (Figure 5) during most epochs (x-axis), thus proving that overfitting isn't present.

As a secondary measure, checking the loss of the model is also incremental in analyzing the model's performance. The loss represents how well the data fit the model. Consistent lowering of the loss is also a good sign. Similar to the concept of accuracy, the best possible sign of loss is if the validation loss is lower than the training loss. As seen in Figure 5, the loss consistently decreased and the validation loss was lower than the training loss.

Numerous measures were also taken to assure a smooth testing process, a process known as hyperparameter tuning. In hyperparameter tuning, one key parameter adjusted is the learning rate. While TensorFlow's in-built tuning mechanisms can help, the best way to improve the model's accuracy is to test random learning rates to see which help. Though often, the last digit of the learning rate is one (ex. 0.001, 0.01).

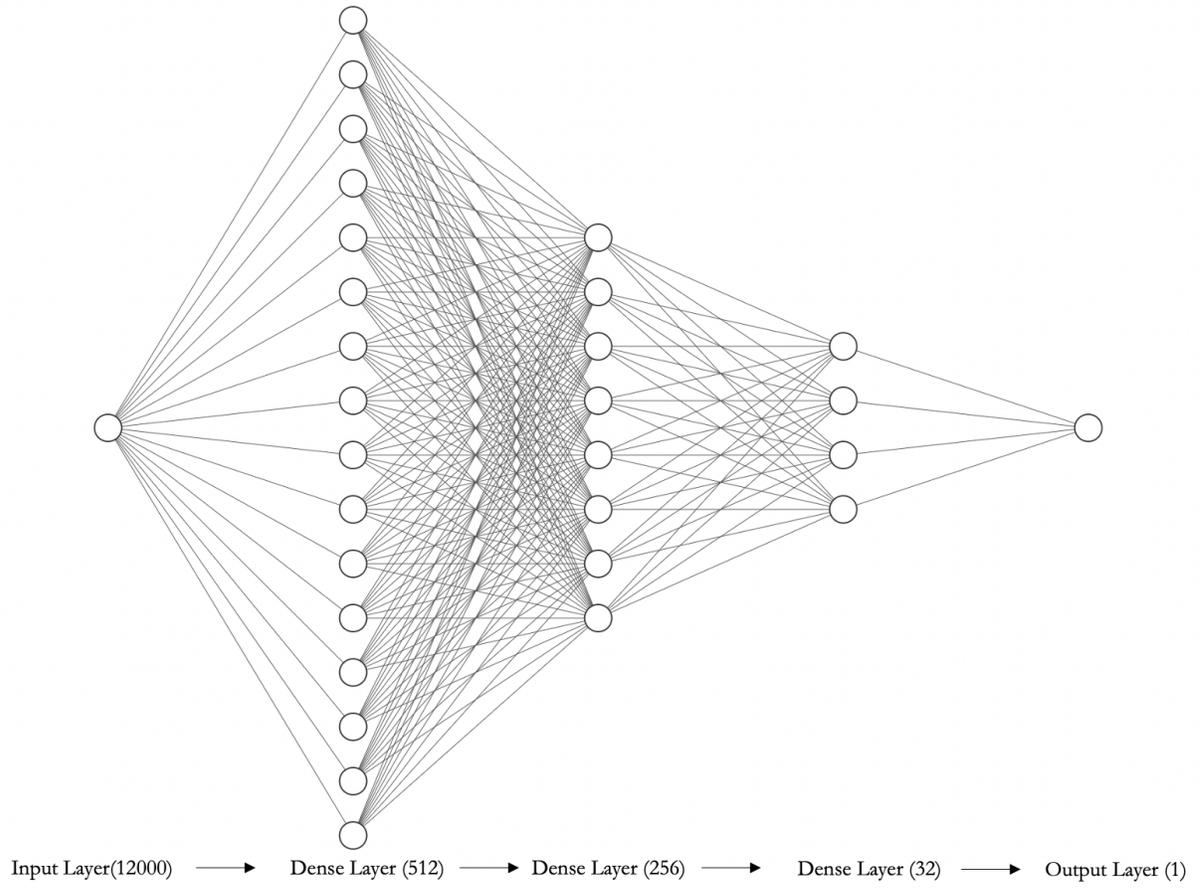

Figure 3: Prototype Neural Network

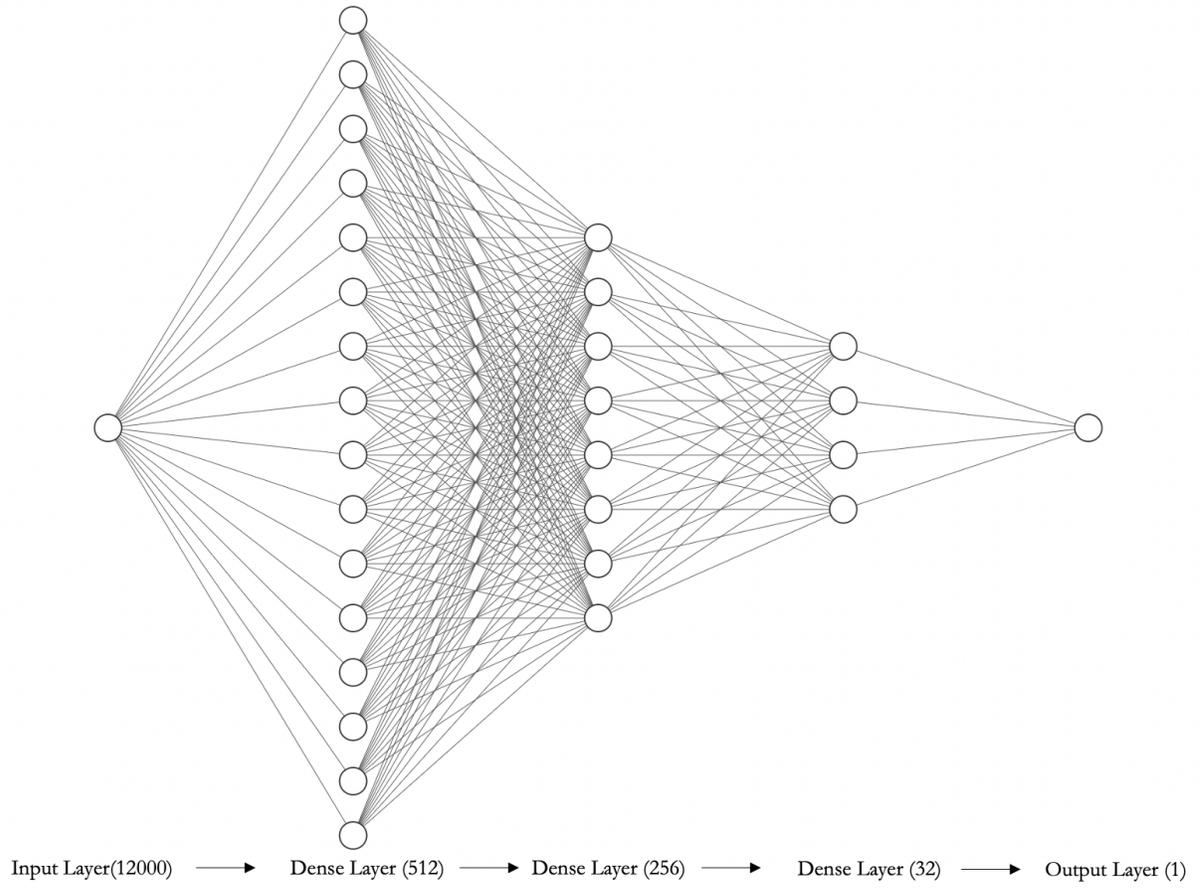

Figure 3: Prototype Neural Network

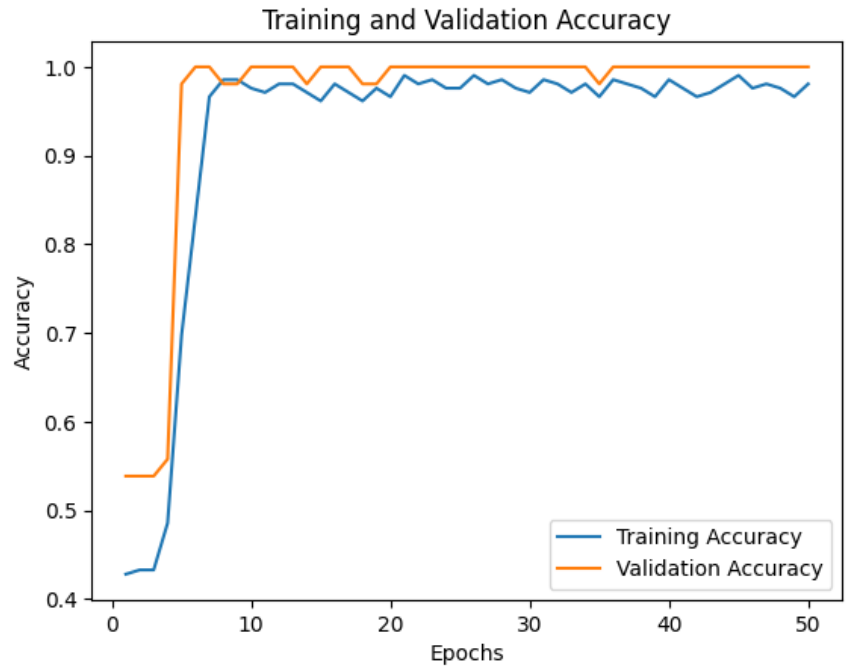

Figure 4: Training and Validation Accuracy

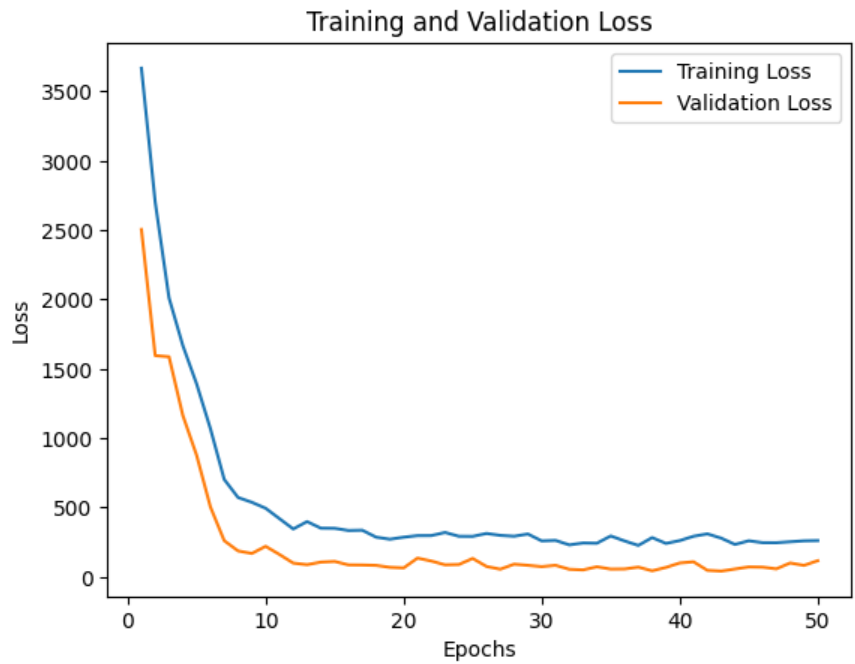

Figure 5: Training and Validation Loss

**Discussion:**

The training accuracy of 98% represents a well-trained model, or neural network, which is suitable for real-world application. The 98% accuracy simply means that the neural network can diagnose the conditions of RVH and RAO/RAE correctly 98% of the time. As discussed earlier, diagnosing the conditions of RVH and RAO/RAE is the best accessible way to ensure that a patient has pulmonary hypertension, as the best publicly available data is from the conditions of RVH and RAO/RAE.

However, training accuracy alone cannot help validate the model. The true test of validation comes with the validation accuracy which is from the testing data. c The validation accuracy of 99% means that the model is ready for real-world use. Lastly, as explained earlier, the loss metrics represent that the model has converged well learning the variations in the data.

As the objective of this project was to create a well-validated neural network model to detect pulmonary hypertension, the goal was achieved. The desired accuracy was 90% but the achieved accuracy was 98%, showing great promise. In addition, the model improves upon current state-of-the-art technology. The neural network model in this project generated a higher accuracy than models from the studies of (Liu et al.) and (Kwon JM, et al.), both of which generated accuracies in the high 80s.

**Conclusion:**

The dense neural network model for ECG with a validation accuracy of around 98% is suitable for real-world applications to detect PH with good accuracy**.** Through early and accurate detection of conditions such as PH, we widen the spectrum of innovation in detecting chronic life-threatening health conditions and reduce associated mortality and morbidity.

Given the limited sample size of only 208 samples, there is a possibility of a lack of variation in the training data which can impact the accuracy of disease detection. Such variations that are missed can pose a serious risk to patients in a real-world scenario as a diagnosis could be missed. To continuously improve the accuracy, the current model will be updated with a Transfer Learning architecture that works best for time-series data. To balance the dataset with proper training samples, additional data will be generated using data augmentation techniques.

The neural network model can also be improved by using a model dedicated to time series data, such as the ECG. Two such techniques known to achieve high accuracy on the ECG are Convolutional Neural Networks (CNN) and LSTM models (Somani et al., 2021). We can combine these models into what is known as a Convolutional LSTM.

The final step would be to create a user-facing application. A model itself cannot be applied in the real world, hence the need for a user application that healthcare workers and patients can use as a screening tool. This application will intake an ECG from the user, run it through the DNN, and lastly give a diagnosis of PH presence. A user-facing front-end application can be created using ReactNative, an open-source UI framework designed by Meta.

**Acknowledgments:**

      I thank my parents for their continuous support through the process of research. I thank Mr. Praveen Kumar, a machine learning scientist, for his mentorship and for taking the time to assist with the optimization of my neural network.

**References:**

Benza, R. L., Miller, D. P., Barst, R. J., Badesch, D. B., Frost, A. E., & McGoon, M. D. (2012). An Evaluation of Long-term Survival From Time of Diagnosis in Pulmonary Arterial Hypertension From the REVEAL Registry. *Chest*, *142*(2), 448–456. https://doi.org/10.1378/chest.11-1460

Benza, R. L., Miller, D. P., Gomberg-Maitland, M., Frantz, R. P., Foreman, A. J., Coffey, C. S., Frost, A., Barst, R. J., Badesch, D. B., Elliott, C. G., Liou, T. G., & McGoon, M. D. (2010). Predicting survival in pulmonary arterial hypertension: insights from the Registry to Evaluate Early and Long-Term Pulmonary Arterial Hypertension Disease Management (REVEAL). *Circulation*, *122*(2), 164–172. https://doi.org/10.1161/CIRCULATIONAHA.109.898122

Butrous, G., Ghofrani, H. A., & Grimminger, F. (2008). Pulmonary vascular disease in the developing world. *Circulation*, *118*(17), 1758–1766. https://doi.org/10.1161/CIRCULATIONAHA.107.727289

Bhattacharya, P. T., & Ellison, M. B. (2022). *Right Ventricular Hypertrophy*. PubMed; StatPearls Publishing. https://www.ncbi.nlm.nih.gov/books/NBK499876/#:~:text=Introduction

CDC. (2019, December 3). *Pulmonary Hypertension | cdc.gov*. Centers for Disease Control and Prevention. https://www.cdc.gov/heartdisease/pulmonary_hypertension.htm#:~:text=What%20is%20pulmonary%20hypertension%3F

Kosaraju 17Cleveland Clinic. (2019, November 21). *Pulmonary Hypertension: Symptoms, Treatment*.

    Cleveland Clinic.

    https://my.clevelandclinic.org/health/diseases/6530-pulmonary-hypertension-ph

Do, E., Boynton, J., Lee, B. S., & Lustgarten, D. (2021). Data Augmentation for 12-Lead

    ECG Beat Classification. *SN Computer Science*, *3*(1).

    https://doi.org/10.1007/s42979-021-00924-x

ECG & Echo Learning. (n.d.). *The ECG leads: electrodes, limb leads, chest (precordial)*

    *leads, 12-Lead ECG (EKG)*. ECG & ECHO.

    https://ecgwaves.com/topic/ekg-ecg-leads-electrodes-systems-limb-chest-precordi

    al/#:~:text=An%20ECG%20lead%20is%20a.

ECG EDU. (n.d.). *Right Ventricular Hypertrophy and the ECG*.

    Https://Www.ecgedu.com/.

    https://www.ecgedu.com/right-ventricular-hypertrophy-ecg/

Gidwani, S., & Nair, A. (2014). The Burden of Pulmonary Hypertension in

    Resource-Limited Settings. *Global Heart*, *9*(3), 297–310.

    https://doi.org/10.1016/j.gheart.2014.08.007

Healio. (n.d.). *Right Ventricular Hypertrophy (RVH) ECG Review*. Www.healio.com.

    https://www.healio.com/cardiology/learn-the-heart/ecg-review/ecg-topic-reviews-

    and-criteria/right-ventricular-hypertrophy-review#:~:text=RVH%20is%20diagnos

    ed%20on%20ECG

Healio (2). (n.d.). *S Wave*. Www.healio.com.

    https://www.healio.com/cardiology/learn-the-heart/ecg-review/ecg-interpretation-t

    utorial/s-wave#:~:text=The%20S%20wave%20is%20the.

Health, S. V. H. (n.d.). *St Vincent's Heart Health*. Www.svhhearthealth.com.au.

    https://www.svhhearthealth.com.au/conditions/pulmonary-hypertension

Kasyanau, A. (2022, July 7). *The role of AI in the future of health care tech*. Fast

    Company.

    https://www.fastcompany.com/90764479/the-role-of-ai-in-the-future-of-health-car

    e-tech

Lau, E. M. T., Humbert, M., & Celermajer, D. S. (2015). Early detection of pulmonary

    arterial hypertension. *Nature Reviews Cardiology*, *12*(3), 143–155.

    https://doi.org/10.1038/nrcardio.2014.191

Mayo Clinic. (2017). *Pulmonary hypertension - Symptoms and causes*. Mayo Clinic.

    https://www.mayoclinic.org/diseases-conditions/pulmonary-hypertension/sympto

    ms-causes/syc-20350697

MedTech Europe. (2020). *The socio-economic impact of AI in healthcare*.

    https://www.medtecheurope.org/wp-content/uploads/2020/10/mte-ai_impact-in-he

    althcare_oct2020_report.pdf

Penn Medicine. (2022). *Pulmonary Hypertension*. Pennmedicine.org.

    https://www.pennmedicine.org/for-patients-and-visitors/patient-information/condi

    tions-treated-a-to-z/pulmonary-hypertension

PHA Staff. (2018). *Types of Pulmonary Hypertension: The WHO Groups - Pulmonary

    Hypertension Association*. Pulmonary Hypertension Association.

    https://phassociation.org/types-pulmonary-hypertension-groups/

Kosaraju 19Ryan, J. J., & Archer, S. L. (2014). The Right Ventricle in Pulmonary Arterial Hypertension. *Circulation Research*, *115*(1), 176–188. https://doi.org/10.1161/circresaha.113.301129

Somani, S., Russak, A. J., Richter, F., Zhao, S., Vaid, A., Chaudhry, F., De Freitas, J. K., Naik, N., Miotto, R., Nadkarni, G. N., Narula, J., Argulian, E., & Glicksberg, B. S. (2021). Deep learning and the electrocardiogram: review of the current state-of-the-art. *EP Europace*. https://doi.org/10.1093/europace/euaa377

The University of Nottingham. (n.d.). *The R Wave - Sinus Rhythm - Normal Function of the Heart - Cardiology Teaching Package - Practice Learning - Division of Nursing - The University of Nottingham*. Www.nottingham.ac.uk. https://www.nottingham.ac.uk/nursing/practice/resources/cardiology/function/r_wave.php#:~:text=Image%3A%20R%20Wave

Vos, T., Lim, S. S., Abbafati, C., Abbas, K. M., Abbasi, M., Abbasifard, M., Abbasi-Kangevari, M., Abbastabar, H., Abd-Allah, F., Abdelalim, A., Abdollahi, M., Abdollahpour, I., Abolhassani, H., Aboyans, V., Abrams, E. M., Abreu, L. G., Abrigo, M. R. M., Abu-Raddad, L. J., Abushouk, A. I., & Acebedo, A. (2020). Global Burden of 369 Diseases and Injuries in 204 Countries and territories, 1990–2019: a Systematic Analysis for the Global Burden of Disease Study 2019. *The Lancet*, *396*(10258), 1204–1222. https://doi.org/10.1016/S0140-6736(20)30925-9

Weatherald, J., & Humbert, M. (2020). The "great wait" for diagnosis in pulmonary arterial hypertension. *Respirology*. https://doi.org/10.1111/resp.13814